\documentclass{article}

\usepackage{microtype}
\usepackage{graphicx}
\usepackage{subfigure}
\usepackage{booktabs} 
\usepackage{textcomp}
\usepackage{epstopdf}
\usepackage{hyperref}
\usepackage[accepted]{icml2021}

\usepackage{icml2021}
\usepackage{color}
\usepackage{multirow}
\usepackage{bm}
\usepackage{amsfonts,amssymb}
\usepackage{amsmath}
\usepackage{graphicx} 
\usepackage{epstopdf}
\usepackage{epsfig}

\newcommand{\xv}[0]{\ensuremath{\boldsymbol{x}} }

\newcommand{\zv}[0]{\ensuremath{\boldsymbol{z}} }

\newcommand{\thetav}[0]{\ensuremath{\boldsymbol{\theta}} }

\newcommand{\epsilonv}[0]{\ensuremath{\boldsymbol{\epsilon}} }
\newcommand{\Omegav}[0]{\ensuremath{\boldsymbol{\Omega}} }

\icmltitlerunning{Sawtooth Factorial Topic Embeddings Guided Gamma Belief Network}

\begin{document}

\twocolumn[
\icmltitle{
Sawtooth Factorial Topic Embeddings Guided Gamma Belief Network
}



\icmlsetsymbol{equal}{*}

\begin{icmlauthorlist}
\icmlauthor{Zhibin Duan}{equal,to}
\icmlauthor{Dongsheng Wang}{equal,to}
\icmlauthor{Bo Chen}{to}
\icmlauthor{Chaojie Wang}{to}\\
\icmlauthor{Wenchao Chen}{to}
\icmlauthor{Yewen Li}{to}
\icmlauthor{Jie Ren}{to}
\icmlauthor{Mingyuan Zhou}{goo}
\end{icmlauthorlist}

\icmlaffiliation{to}{National Laboratory of Radar Signal Processing, Xidian University, Xi’an, China.}
\icmlaffiliation{goo}{McCombs School of Business The University of Texas at Austin, Austin, TX 78712, USA}

\icmlcorrespondingauthor{Bo Chen}{bchen@mail.xidian.edu.com}

\icmlkeywords{Machine Learning, ICML}

\vskip 0.3in
]



\printAffiliationsAndNotice{\icmlEqualContribution} 
   
\begin{abstract}



Hierarchical topic models such as the gamma  belief network (GBN) have delivered promising results in mining multi-layer document representations and discovering interpretable topic taxonomies. However, they often assume in the prior that the topics at each layer are independently drawn from the Dirichlet distribution, ignoring the dependencies between the topics both at the same layer and across different layers. To relax this assumption, we propose sawtooth factorial topic embedding guided GBN, a deep generative model of documents that captures the dependencies and semantic similarities between the topics in the embedding space. Specifically, both the words and topics are represented as embedding vectors of the same dimension. The topic matrix at a layer is factorized into the product of a factor loading matrix and a topic embedding matrix, the transpose of which is set as the factor loading matrix of the layer above. Repeating this particular type of factorization, which shares components between adjacent layers, leads to a structure referred to as sawtooth factorization. An auto-encoding variational inference network is constructed to optimize the model parameter via stochastic gradient descent. Experiments  on big corpora show that our models outperform other neural topic models on extracting deeper interpretable topics and deriving better document representations.

\end{abstract}

\section{Introduction}


Probabilistic topic models, such as latent Dirichlet allocation (LDA) \cite{blei2003latent} and Poisson factor analysis (PFA) \cite{zhou2012beta},  have the ability to discover the underlying semantic themes from a collection of documents, achieving great success in text analysis. 
In general, a topic model represents each document as a mixture of latent topics, each of which describes an interpretable semantic concept. While being widely used, vanilla topic models assume that topics are independent and there are no structures among them, which limits those models’ ability to explore any hierarchical thematic structures. 
To remove its limitation, a series of  hierarchical extensions, including nonparametric Bayesian hierarchical prior based topic models \cite{blei2010nested, paisley2014nested}, deep PFA \cite{gan2015scalable, henao2015deep}, gamma belief network (GBN) \cite{zhou2015poisson, cong2017deep}, and Dirichlet belief network (DirBN) \cite{zhao2018dirichlet}, have been proposed.
Commonly, these models learn directed acyclic graph (DAG)-structured hierarchical topics, which assumes that the topics in the upper layers are more general/abstract than those in the lower layers. 
Consequently,  revealing hierarchical relations between topics provides the user an intuitive 
way to better understand text data.






With the development of deep neural networks (DNN), there is a growing interest in developing neural topic models (NTMs). Specifically, most neural topic models are based on variational auto-encoders (VAEs) \cite{kingma2013auto, rezende2014stochastic}, which employ a variational inference network (encoder) to approximate the posterior distribution and are equipped with a decoder to reconstruct the document's Bag-of-Words (BOW) representation \cite{miao2016neural, srivastava2017autoencoding, card2017neural}.
However, most NTMs rely on Gaussian latent variables, which often fail to well approximate the posterior distributions of sparse and nonnegative document latent representations.
To address this limition, \citet{zhang2018whai} develop Weibull hybrid autoencoding inference (WHAI) for deep LDA, which infers posterior samples via a hybrid of stochastic-gradient MCMC and autoencoding variational Bayes.
As a hierarchical neural topic model, WHAI shows attractive qualities in multi-layer document representation learning and hierarchical explainable topic discovery. 
Compared with traditional Bayesian probabilistic topic models, these NTMs usually enjoy better flexibility and scalability, which are important for modeling large-scale data and performing downstream tasks \cite{zhang2019variational, wang2020learning, NEURIPS2020_26178fc7, wang2020deep, duan2021enslm, zhao2021topic}.

\begin{figure}[!t]
 \centering
  \subfigure[]{\includegraphics[width=9.5mm]{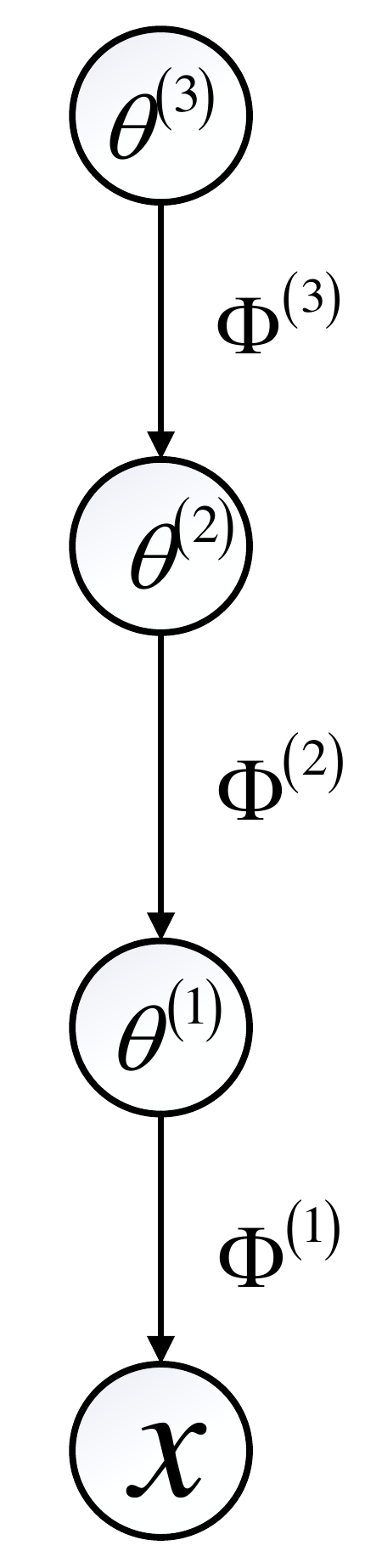}
  \label{generate}}
  \quad
  \subfigure[]{\includegraphics[width=58.5mm]{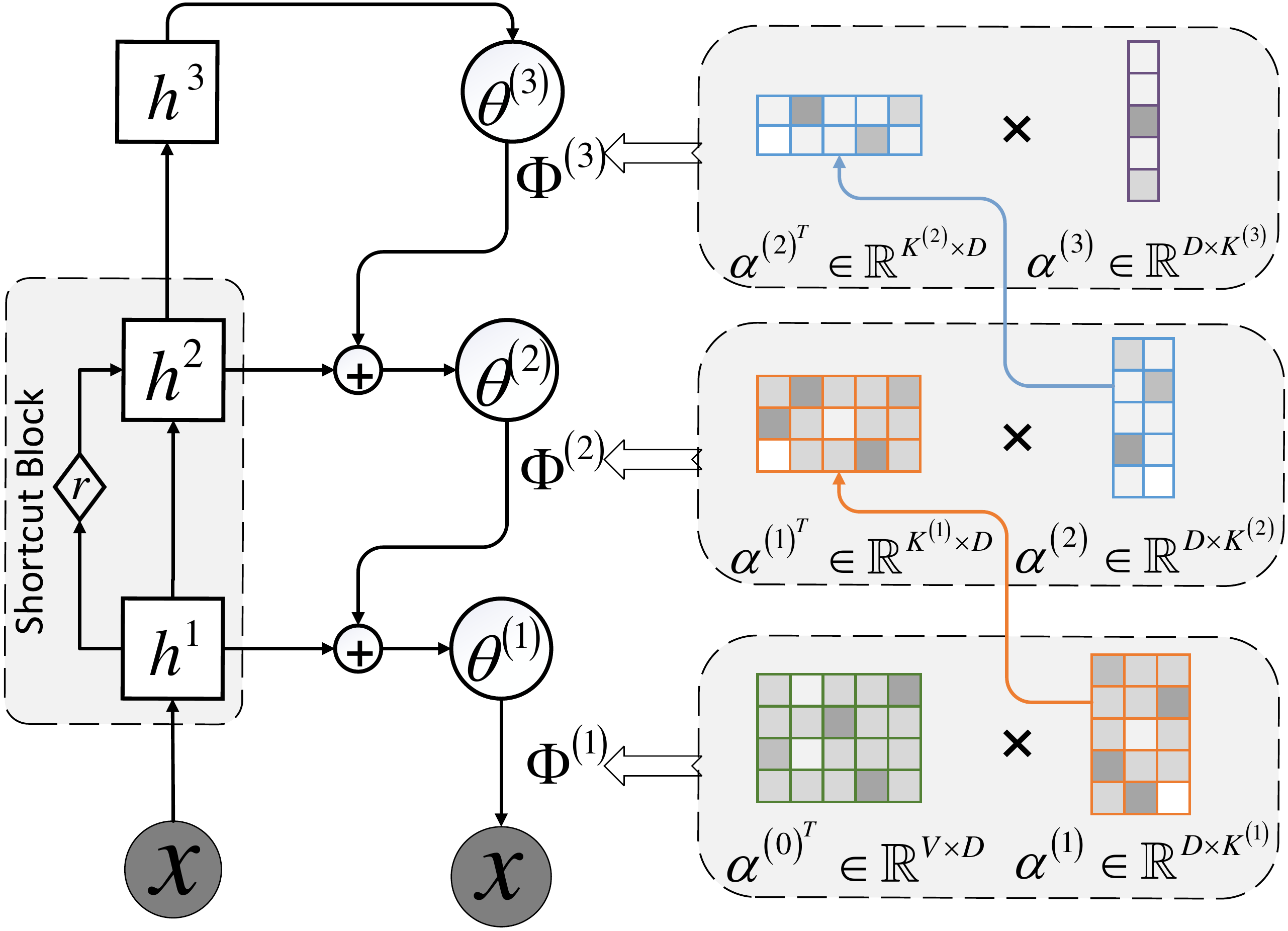}
  \label{fig_model_inference}}
  \caption{\small (a) Gamma belief network and (b) overview of the proposed SawETM and its corresponding hierarchical upward and downward encoder networks }
 \label{model}
\end{figure}

Despite their attractive performance, existing hierarchical topic models such as GBN often assume in the prior that the topics at each layer are independently drawn from the Dirichlet distribution, ignoring the dependencies between the topics both at the same layer and across different layers.
To relax this assumption, we propose the Sawtooth Factorial {T}opic {E}mbeddings Guided GBN (SawETM), a deep generative model of documents that captures the dependencies and semantic similarities between the topics in the embedding space.
Specifically, as sketched in Fig. \ref{model}, both the words and hierarchical topics are first converted into the shared embedding space. Then we develop the Sawtooth Connection technique to capture the dependencies between the topics at different layers, where the factor loading at layer $l$ is the factor score at layer $l-1$, which enables the hierarchical topics to be coupled together across all layers. 
Our work is inspired by both GBN \cite{zhou2015poisson}, a multi-stochastic-layer hierarchical topic model,  and  the embedding topic models \cite{dieng2019dynamic,dieng2020topic}, which represent the words and single layer topic as embedding vectors.
The proposed Sawtooth Connector is a novel method that combines the advantages of both models for hierarchical topic modeling.


We further note that previous work on NTMs has been restricted to shallow models with one or three layers of stochastic latent variables, which could limit their ability. 
Generally, due to the well-known component collapsing problem of VAEs \cite{sonderby2016ladder}, constructing a deep latent variable model is challenging work. 
As discussed in \citet{child2020very}, the hierarchical VAEs with a sufficient depth can not only learn arbitrary orderings over observed variables but also learn more effective latent variable distributions, if such distributions exist. 
Moving beyond text modeling, the recent development on image generation has shown its promising performance and outstanding generation ability \citep{maaloe2019biva, vahdat2020nvae, child2020very}. Inspire by their work, we carefully design the inference network of SawETM in a deep hierarchical VAE framework to improve the model's ability of modeling textual data. In particular, we propose the integration of a skip-connected deterministic upward path 
and a stochastic path to approximate the posterior of the latent variables and obtain hierarchical representations of a document. We also provide customized training strategies to build deeper neural topic models. To the best of our knowledge, SawETM is the first neural topic model that well supports a deep network structure (e.g., 15).
Our main contributions are summarized as follows:
\begin{itemize}
\setlength{\itemsep}{3pt}
\setlength{\parsep}{0pt}
\setlength{\parskip}{0pt}

\item To move beyond the independence assumption between the topics of two adjacent layers in most hierarchical topic models, the Sawtooth Connection technique is developed to extend 
GBN by capturing the dependencies and semantic similarities between the topics in the embedding space.

\item To avoid posterior collapse, We carefully design a residual upward-downward inference network for SawETM to improve the model's ability of modeling count data and  approximating sparse, non-negative and skewed document latent representations.

\item Overall, SawETM, a novel hierarchical NTM 
equipped with a flexible training algorithm, is proposed to infer multi-layer document representations and discover topic hierarchies in both the embedding space and vocabulary space. Experiments on big corpora show that our models outperform other NTMs on extracting deeper interpretable topics and deriving better document representation.
\end{itemize}

\section{Related work}
The proposed model in this paper marries a hierarchical neural topic model with word embedding, resulting in a deep generative framework for text modeling. The related work can be roughly divided into two categories, one is the research on constructing neural topic model and the other is on leveraging word embedding for topic models.

\paragraph{Neural topic models}
Most existing NTMs can be regarded as an extension of Bayesian topic models like LDA within the VAE framework for text modeling, where the latent variables $\zv$ can be viewed as topic proportions.
NTMs usually utilize a single-layer network as their decoder, e.g., $\phi(\zv) = \mbox{softmax}(\bm{W}\zv)$ \cite{srivastava2017autoencoding}, where $\bm{W}$ is a learnable weights between topics and words. 
Different NTMs may place different distributions on latent variable $\zv$, such as Gaussian and Dirichlet distributions \cite{miao2016neural, burkhardt2019decoupling, nan2019topic, wang2020neural}. 
Different from these models that only focus on a single layer latent variable model, \citet{zhang2018whai} propose WHAI, a hierarchical neural topic model that employs a Weibull upward-downward variational encoder to infer multi-layer document latent representations and use GBN as a decoder. All of these works focus on relatively shallow models with one or three layers of stochastic latent variables.
We note that our work is based on WHAI, but different from it, we propose a new decoder to capture the dependencies of topics and a new powerful encoder to approximate the posteriors, which result in a deeper neural topic model.

\paragraph{Word embedding topic models} Word embeddings can capture word semantics at low-dimensional continuous space and are well studied in neural language models \cite{bengio2003neural, mikolov2013efficient, mikolov2013distributed, levy2014neural}. Due to their ability to capture semantic information, there is a growing interest in applying word embeddings to topic models. 
Pre-trained word embeddings, such as GloVe \cite{pennington2014glove} and word2vec \cite{mikolov2013distributed}, can serve as complementary information to guide topic discovery, which is effective to alleviate the sparsity issue in topic models \cite{zhao2017word, zhao2018inter, li2016topic}.
\citet{dieng2020topic} propose an embedding-based topic model (ETM), which directly models the similarity in its generative process, rather than via a Dirichlet distribution.
What should be noted is that our model is related to ETM 
in modeling the correlation between words and topics using their semantic similarities. 
Different from ETM, the Sawtooth Connection can be seen as injecting the learned knowledge information of a lower layer to a higher layer, which alleviates the sparsity issue in the higher layer.

\section{The proposed model}

%
In this section, we develop SawETM for text analysis, which aims at mining document's multi-layer representations and exploring topic hierarchies. 
The motivation for designing SawETM focuses on tackling two main challenges: $(i)$ a hierarchical decoder to construct the dependencies between topics; $(ii)$ designing expressive neural networks to approximate the  posterior distribution more accurately.
Below, we will first describe the decoder and encoder of SawETM, and then discuss the model's properties. Finally, we will provide details of model inference and stable training techniques. 



\subsection{Document decoder: Sawtooth Factorial Topic Embeddings Guided GBN} 
To explore hierarchical thematic structure from a collection of documents, SawETM  adapts GBN of \citet{zhou2015poisson} as its generative module (decoder).  Different from GBN, where the interactions between the topics of two adjacent layers are modeled as a Dirichlet distribution, SawETM utilizes the Sawtooth Connection (SC) technique to couple hierarchical topics across all layers in the shared embedding space. Specifically, assuming the observations are multivariate count vectors $\xv_j \in \mathbb{Z}^{K_0}$, 
the generative model of SawETM with $L$ layers, from top to bottom, can be expressed as
\begin{align} \label{SawETM}
	&\bm{\theta}_j^{(L)} \sim \mbox{Gam}(r, c_j^{(L + 1)}), \notag \\
	&\bm{\theta}_j^{(l)} \sim \mbox{Gam}(\bm{\Phi}^{(l + 1)} \bm{\theta}_j^{(l + 1)}, c_j^{(l + 1)}), \quad l=1, \cdots, L-1 \notag\\
	&\bm{\Phi}^{(l)}_k = \mbox{Softmax}({\bm{\alpha}^{(l-1)}}^{T} \bm{\alpha}^{(l)}_k), \quad l=1, \cdots, L \notag \\
	&\bm{x} _j \sim \mbox{Pois}(\bm{\Phi} ^{(1)} \bm{\theta} _j ^{(1)})
\end{align}
where, the count vector $\xv_j$ (e.g., the bag-of-word of document $j$) is factorized as the product of the factor loading matrix $\bm{\Phi}^{(1)}$ (topics), and gamma distributed factor scores $\bm{\theta} _j^{(1)}$ (topic proportions), under the Poisson likelihood; and the hidden units $\bm{\theta} _j^{(l)} \in \mathbb{R}^{K_l}_+$ of layer $l$ are further factorized into the product of the factor loading $\bm{\Phi} ^{(l+1)} \in \mathbb{R}_+^{K_{l} \times K_{l+1}}$ and hidden units of the next layer to infer a multi-layer latent representations. $K_l$ denotes the topic numbers at layer $l$.
the vector $\bm{\alpha}^{(l)}_k \in \mathbb{R}^{D}$ is a distributed representation of the $k^{th}$ topic at layer $l$ in the semantic space of words. Especially, $\bm{\alpha} ^{(0)} \in \mathbb{R}^{D \times V}$ is the word embedding matrix. 
$\bm{\Phi}^{(l)}_k$ captures the relationship between topics of two adjacent layers and is calculated by the SC technique. In detail, SC first calculates the semantic similarities between the topics of two adjacent layers by the inner product of their embedding vectors and then applies softmax normalization to make sure the sum of each column of $\bm{\Phi} ^{(l)}$ is equal to one. Note that, the factor loading at layer $l$ is the factor score at layer $l-1$, which constructs the dependency between the parameters of the adjacent layers. Repeating this process, the hierarchical topic can be coupled together across all layers. 

To verify the role of SC, we also consider two simple variants for ablation study,  formulated as
\begin{align}
    &\bm{\Phi}^{(l)}_k = \mbox{Softmax}(\bm{W}_k^{(l)}), \quad l=1, \cdots L \label{eq_w}  \\
    &\bm{\Phi}^{(l)}_k = \mbox{Softmax}(\bm{\alpha} ^{(l)} \bm{\beta}_k^{(l)}), \quad l=1,\cdots, L \label{eq_unshare} 
\end{align} 
where, Eq.~\eqref{eq_w} directly models the dependence via the learnable parameters $\bm{W}$, which is a common choice in most NTMs. Eq.~\eqref{eq_unshare} employs a similar decomposition form as SC but does not share topic embeddings between two adjacent layers. We refer to the first variant as a deep neural topic model (DNTM), and the second as a deep embedding topic model (DETM). Note that when the number of layers is set to one, the DETM has the same structure as SawETM.


\subsection{Document encoder: upward and downward encoder networks} 
The goal of encoder is to reparameterize the variational posterior distribution $q(\bm{\theta}_j)$, which denotes topic proportions. While the gamma distribution, satisfying the nonnegative constraint and encouraging sparsity, appears to be a natural choice, it is not reparameterizable and therefore not amenable to gradient
descent based optimization. Here we consider Weibull distribution because $i)$, latent variable $x\sim \mbox{Weibull}(k,\lambda )$ can be easily reparameterized as:
\begin{equation}
x = \lambda {( - \ln (1 - \varepsilon ))^{1/k}},~~\varepsilon \sim \mbox{Uniform}(0,1).
\label{eq_reparameterize}
\end{equation}
$ii)$, The KL divergence from the gamma to Weibull distributions has an analytic expression \cite{zhang2018whai}:
\begin{equation} \notag
\setlength{\abovedisplayskip}{3pt}
\setlength{\belowdisplayskip}{3pt}
\begin{aligned}
KL(\mbox{Weibull}(k,\lambda)||\mbox{Gamma}(\alpha,\beta)) = \frac{\gamma\alpha}{k} - \alpha \mbox{log}\lambda+\\
\mbox{log}k + \beta\lambda\Gamma(1+\frac{1}{k}) - \gamma -1 -\alpha \mbox{log}(\beta) + \mbox{log}\Gamma(\alpha).
\end{aligned}
\end{equation} 
where $\gamma$ is the Euler-Mascheroni constant.
What's more, designing a capacious inference network is necessary for deep VAEs training. Inspired by the series of studies about deep generative models \cite{sonderby2016ladder,zhang2018whai, vahdat2020nvae, child2020very}, we develop an upward-downward inference network, which contains a bottom-up residual deterministic path and a top-down stochastic path.

\paragraph{Upward-downward inference model: }
Like most neural topic models, the exact posterior distribution for $\bm{\theta}^{(l)}$ is intractable and needs to be approximated. Following VAEs \cite{kingma2013auto, rezende2014stochastic}, we define the variational distribution as $q(\thetav_j|\xv_j)$, which need to be flexible enough to well approximate the true posterior distribution. 
The variational distribution is factorized with a bottom-up structure as 
\begin{equation} \label{q_theta}
	\begin{aligned}
	q(\thetav_j|\xv_j)=q(\thetav_j^{(1)}|\xv_j) \prod_{l=1}^{L-1} q(\thetav_j^{(l+1)}|\thetav_j^{(l)})
	\end{aligned}
\end{equation}
Here we emphasize that this hierarchical structure makes the top stochastic latent variables tend to collapse into the prior, especially when the layer is large \cite{sonderby2016ladder}.
To address this problem, SawETM first parameters a skip-connected deterministic upward path to obtain the latent representations of input $\xv$:
\begin{equation} \notag
	\begin{aligned}
	&\bm{h}_j^{(l)} = \bm{h}_j^{(l-1)} + \mbox{MLP}(\bm{h}_j^{(l-1)}) \\
	&\hat{\bm{k}}_j^{(l)} = \mbox{Relu}(\mbox{Linear}(\bm{h}_j^{(l)})) \\
	&\hat{\bm{\lambda}} _j^{(l)}= \mbox{Relu}(\mbox{Linear}(\bm{h}_j^{(l)}))
	\end{aligned}
\end{equation}
where $\bm{h}^{(0)} = \xv$, $\mbox{MLP}$ is a two layer fully connected network, $\mbox{Linear}$ is a single linear layer, and $\mbox{Relu}$ applies nonlinear activation function. 
SawETM combines the obtained latent features with the prior from the stochastic up-down path to 
construct the variational posterior:
\begin{align}
	&q(\bm{\theta}_j^{(l)} | \bm{\Phi}^{(l+1)},\bm{h}_j^{(l)},\bm{\theta}_j^{(l+1)}) = 
	\mbox{Weibull}(\bm{k}_j^{(l)}, \bm{\lambda}_j^{(l)}) \label{q_weibull} \\
	&\bm{k}_j^{(l)} = \mbox{Softplus}(\mbox{Linear}(\bm{\Phi}^{(l+1)} \bm{\theta}_j^{(l+1)} \oplus \hat{\bm{k}}_j^{(l)})) \notag \\
	&\bm{\lambda}_j^{(l)} = \mbox{Softplus}(\mbox{Linear}(\bm{\Phi}^{(l+1)} \bm{\theta}_j^{(l+1)} \oplus \hat{\bm{\lambda}}_j^{(l)})) \notag
\end{align}
where $\oplus$ denotes the concatenation at topic dimension, and $\mbox{Softplus}$ applies $\mbox{log}(1+\mbox{exp}(\cdot))$ nonlinearity to each element to ensure positive Weibull shape and scale parameters. The Weibull distribution is used to approximate the gamma-distributed conditional posterior, and its parameters $\bm{k}_j^{(l)} \in \mbox{R}^{K_l}$ and $\bm{\lambda}_j^{(l)} \in \mbox{R}^{K_l}$ are both inferred by combining the bottom-up likelihood information and the prior information from the generative distribution using the neural networks. The inference network is structured as
\begin{align}\label{inference network}
q(\bm{\theta}|x)=q(\bm{\theta}^{(L)}|x) \prod_{l=1}^{L-1} q(\bm{\theta}^{(l)}|\bm{\theta}^{(l+1)},x).
\end{align}
compared with Eq.~\eqref{q_theta}, the residual upward pass in SawETM allows all the latent variables to have a deterministic dependency on input $x$, thus the top stochastic latent variables could receive efficient information and will be empirically less likely to collapse. Note that, compared with the inference network in WHAI \cite{zhang2018whai}, we construct a more powerful residual network structure to better approximate the true posteriors. 

\subsection{Model properties}
SawETM inherits the good properties of both deep topic model and word embeddings, as described below.

\paragraph{Semantic topic taxonomies:} The loading matrices  $\bm{\Phi}^{(l)}$ in Eq.~\eqref{SawETM} 
capture the semantic correlations of the topics of adjacent layers. Using the law of total expectation, we have
\begin{equation} \label{expectation of x}
	\mathbb{E}\left[ \bm{x}_j | \bm{\theta}_j^{(l)} ,\left\{ \bm{\Phi}^{(t)}, c_j^{(t)} \right\}_{t=1}^{l} \right] = \left[ \prod_{t=1}^l \bm{\Phi}^{(t)} \right] \frac{\bm{\theta}_j^{(l)}}{\prod_{t=2}^l c_j^{(t)}}.
\end{equation}
Therefore, $\left[\prod_{t=1}^{l-1} \bm{\Phi}^{(t)}\right] \bm{\Phi}^{l}$ is naturally interpreted as the projection of topic $\bm{\Phi}^{(l)}$ to the vocabulary space, providing us with a way to visualize the topics at different semantic layers. The topics in the bottom layers are more specific and become increasingly more general when moving upward, as shown in Fig.\ref{fig:hier_structure}.  

\paragraph{Hierarchical topics in the same embedding space:}
In SawETM, both words and hierarchical topics are represented with embedding vector (e.g., $\alpha^{(l)} \in \mathbb{R}^{D \times K_l}$), and the topic can be defined by the Sawtoorh Connection (e.g. the  $l ^{\text{th}}$ layer topic is defined as $\bm{\Phi}^{(l)}=\mbox{softmax}({\bm{\alpha}^{(l-1)}}^{T} \bm{\alpha}^{(l)})$). The first advantage is that the learned words and hierarchical topics can be projected into the same embeding space, which is shown in Fig. \ref{figure:embedding}.
And the second advantage is that SawETM can establish dependencies between different layers, which can be seen as the learned knowledge information of a lower layer can be injected into a higher layer. The intuition is there is semantic relation between the same layer topics, such as the topic about `basketball' have a strong relation with the topic about `game', which should be considered by the higher layer topics. Note that, other hierarchical topic models such as GBN usually assume the hierarchical topic is independent and ignore this semantic structure \cite{blei2007correlated}, while SawETM try to capture this structure in the embedding space.



\subsection{Inference and estimation} \label{Inference and Estimation}
Similar to VAEs, the training objective of SawETM is the maximization of an Evidence Lower Bound (ELBO):
\begin{equation} \label{ELBO}
\setlength{\abovedisplayskip}{3pt}
\setlength{\belowdisplayskip}{3pt}
\begin{aligned}
	\mathcal{L} &= \sum_{j=1}^{J} \mbox{E}_{q(\bm{\theta}|\xv_j)} [\mbox{ln} p(\xv_j|\bm{\Phi}^{(1)},\bm{\theta}_j^{(1)})]\\
	&- \sum_{j=1}^{J} \sum_{l=1}^{L} \mbox{KL}(q(\bm{\theta}^{(l)}_j) || p(\bm{\theta}^{(l)}_j | \bm{\Phi}^{(l+1)},\bm{\theta}^{(l+1)}_j)))
\end{aligned}
\end{equation} 
where, $\bm{\Phi^{(l)}} = \mbox{softmax}({\bm{\alpha^{(l-1)}}}^{T} \bm{\alpha^{(l)}})$, $q(\bm{\theta}^{(l)}_j)$ is the Weibull variational distribution in Eq.~\eqref{q_weibull}, and  $p(\bm{\theta}^{(l)}_j)$ is the gamma prior  distribution in Eq.~\eqref{SawETM}. The first term is the expected log-likelihood or reconstruction error, while the second term is the Kullback--Leibler (KL) divergence that constrains $q(\bm{\theta}^{(l)}_j)$ to be close to its prior $p(\bm{\theta}^{(l)}_j)$ in the generative model. Thanks to the analytic KL expression and easy reparameterization of the Weibull distribution, the gradient of the ELBO with respect to $\{\bm{\alpha^{(l)}}\}^{L}_{(l=0)}$ and other parameters in the inference network can be accurately evaluated. As describe in Algorithm.~\ref{alg:example}, the encoder parameters $\Omegav $ and decoder parameters $\bm{\Psi}$ in SawETM are updated by SGD, which makes faster inference at both train and test stages compared to Gibbs Sampling. This also helps differ SawETM from WHAI, which updates the global parameters by SG-MCMC and is limited to update local and global parameters alternately.

\begin{algorithm}[tb]
   \caption{Upward-Downwar Autoencoding Variational Inference for SawETM }
   \label{alg:example}
\begin{algorithmic}
   \STATE Set mini-batch size $m$ and the number of layer $L$
   \STATE Initialize the encoder parameters $\Omegav $ and decoder parameters $\bm{\Psi}$;
   \FOR{$\text{iter = 1,2,} \cdot \cdot \cdot $} 
   \STATE Randomly select a mini-batch of $m$ documents to form a subset ${\rm{X}} = {\left\{ {{\xv_i}} \right\}_{1,m}}$;
   \STATE Dram random noise $\left\{ {\bm{\varepsilon} _i^l} \right\}_{i = 1,l = 1}^{m,L}$ from uniform distribution;
    \STATE Calculate  $\nabla {}_{\Omegav , {\bm{\Psi}}}L\left( {\Omegav ,{\bm{\Psi}};{\rm{X,}} \left\{ {{\bm{\varepsilon} _i^l}} \right\}_{i = 1,l = 1}^{m,L}} \right)$ according to Eq.~\eqref{ELBO}, and update encoder parameters $\Omegav $ and decoder parameter $\bm{\Psi}$ jointly ;
   \ENDFOR
\end{algorithmic}
\end{algorithm} 


\subsection{Stable training} 
It is a great optimization challenge to train a deep hierarchical VAE in practice, due to the well-known posterior collapse and unbounded KL divergence in the objective \cite{razavi2019preventing,child2020very}. Here, we propose three approaches for stabilizing the training. We emphasize that all these approaches are applied in WHAI and other hierarchical neural topic model variants for fair comparison when we perform experiments.

\paragraph{Shape parameter skipping of Weibull distribution:} 
As shown in Eq.~\eqref{eq_reparameterize}, when the sampled noise $\epsilonv$ is close to 1, e.g., $0.98$, and the Weibull shape parameter $k$ is less than $\text{1e-3}$, the $x$ will be extremely huge, which could destabilize the training process. In practice, we constrain the shape parameter $k$ such that $k \ge 0.1$ to avoid extreme value. A similar setting can be found in \citet{fan2020bayesian}, who view $k$ as a hyperparameter and set it as $0.1$.
\paragraph{Warm-up:} The variational training criterion in Eq.~\eqref{ELBO} contains the likelihood term $p(\xv_j|\bm{\Phi}^{(1)},\bm{\theta}_j^{(1)})$ and the variational regularization term. During the early training stage, 
the variational regularization term causes some of the latent units to become inactive before their learning useful representation \cite{sonderby2016ladder}.
We solve this problem by first training the parameters only using the reconstruction error,and then adding the KL loss gradually with a temperature coefficient:
\begin{align}
	\mathcal{L} &= \sum_{j=1}^{J} \mbox{E}_{q(\bm{\theta}|\xv_j)} [\mbox{ln}p(\xv_j|\bm{\Phi}^{(1)},\bm{\theta}_j^{(1)})] \notag \\
	&- \beta \sum_{j=1}^{J} \sum_{l=1}^{L} \mbox{KL}(q(\bm{\theta}^{(l)}_j) || p(\bm{\theta}^{(l)}_j | \bm{\Phi}^{(l+1)},\bm{\theta}^{(l+1)}_j)))
	\label{ELBO}
\end{align}
where $\beta$ is increased from 0 to 1 during the first N training epochs. This idea has been considered in previous VAE-based algorithms \cite{raiko2007building,higgins2016beta}. 
\paragraph{Gradient clipping:}Optimizing the unbounded KL loss often causes the sharp gradient during training \cite{child2020very}, we address this by clipping gradient with a large L2-norm above a certain threshold, which we set 20 in all experiments. This technique can be easily implemented and allows networks to train smoothly.

\begin{figure*}[!ht]
\centering
\subfigure[20NG]{
\includegraphics[width=0.32\linewidth]{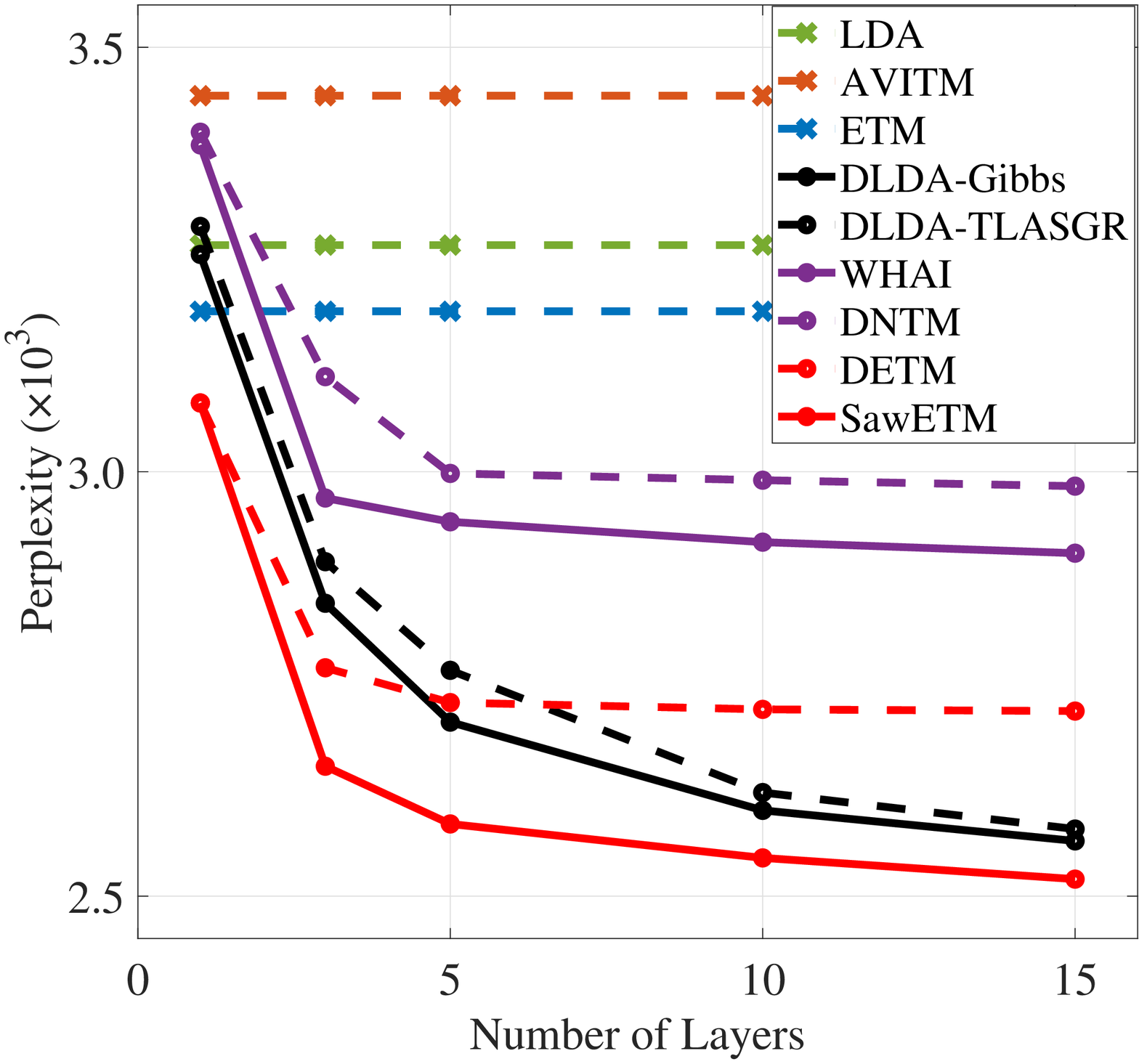}
} \hspace{-5mm}
\quad 
\subfigure[RCV1]{
\includegraphics[width=0.32\linewidth]{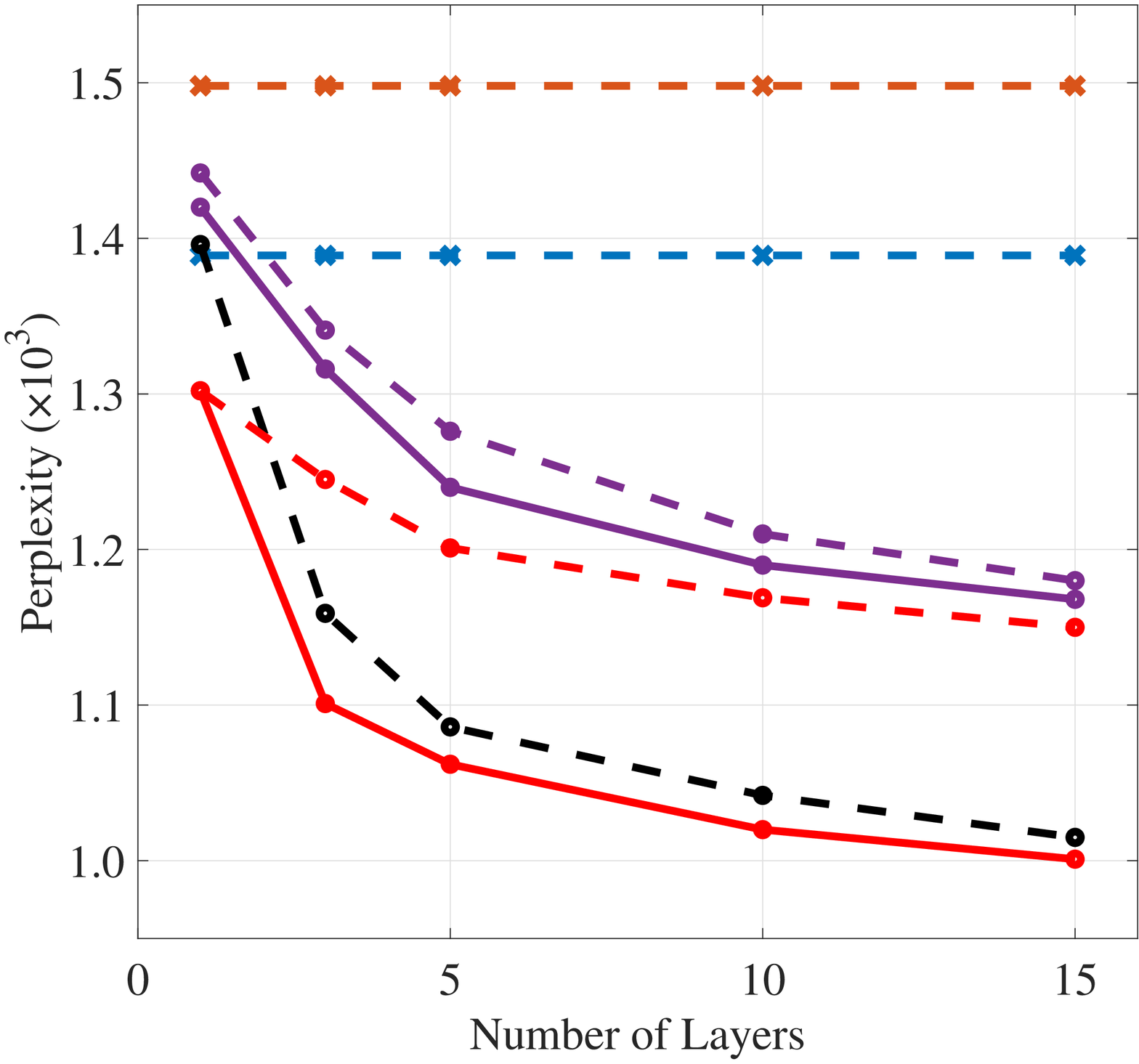}
} \hspace{-5mm}
\quad 
\subfigure[PG19]{
\includegraphics[width=0.32\linewidth]{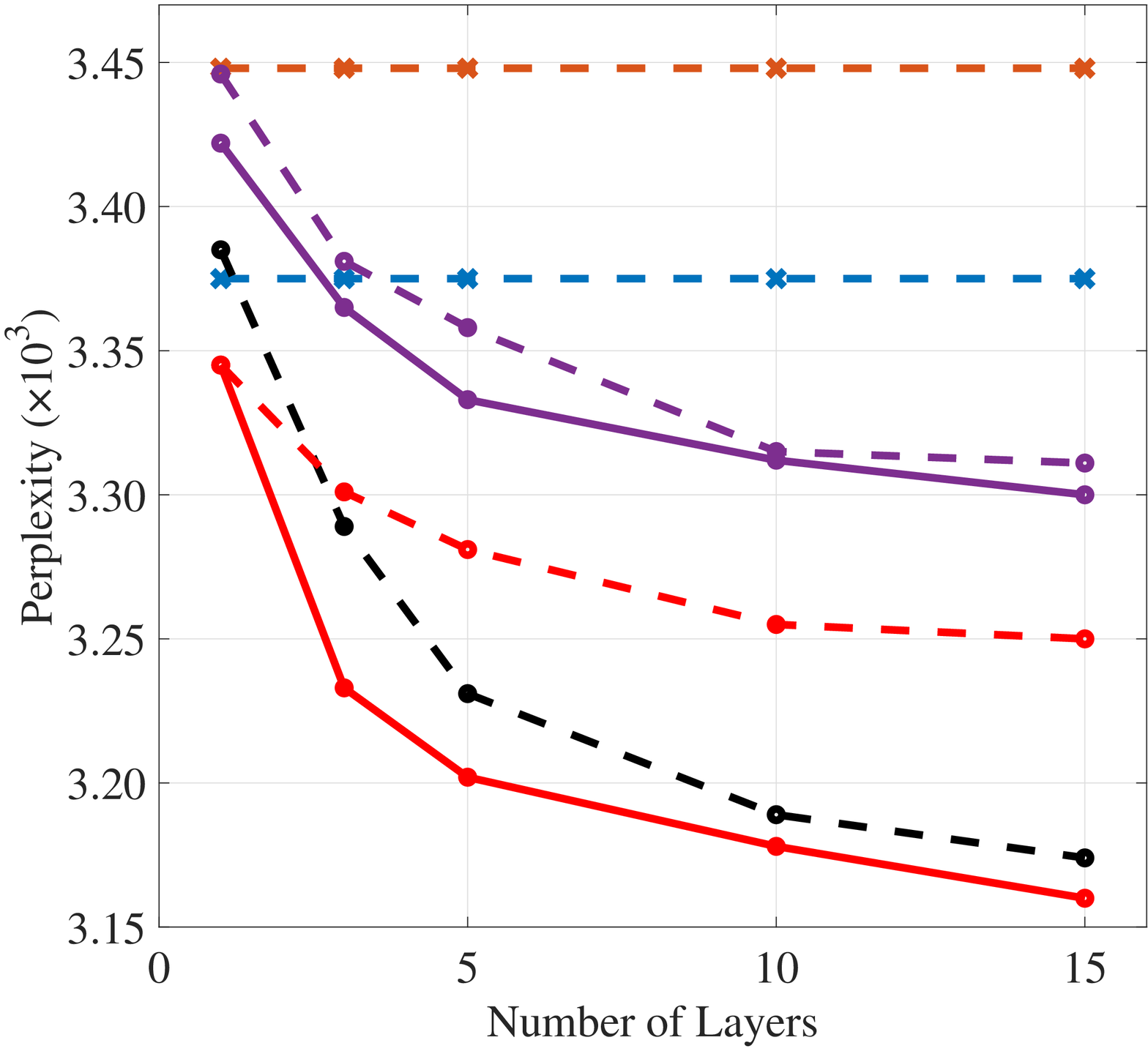}
}
\quad 
\subfigure[20NG]{
\includegraphics[width=0.32\linewidth]{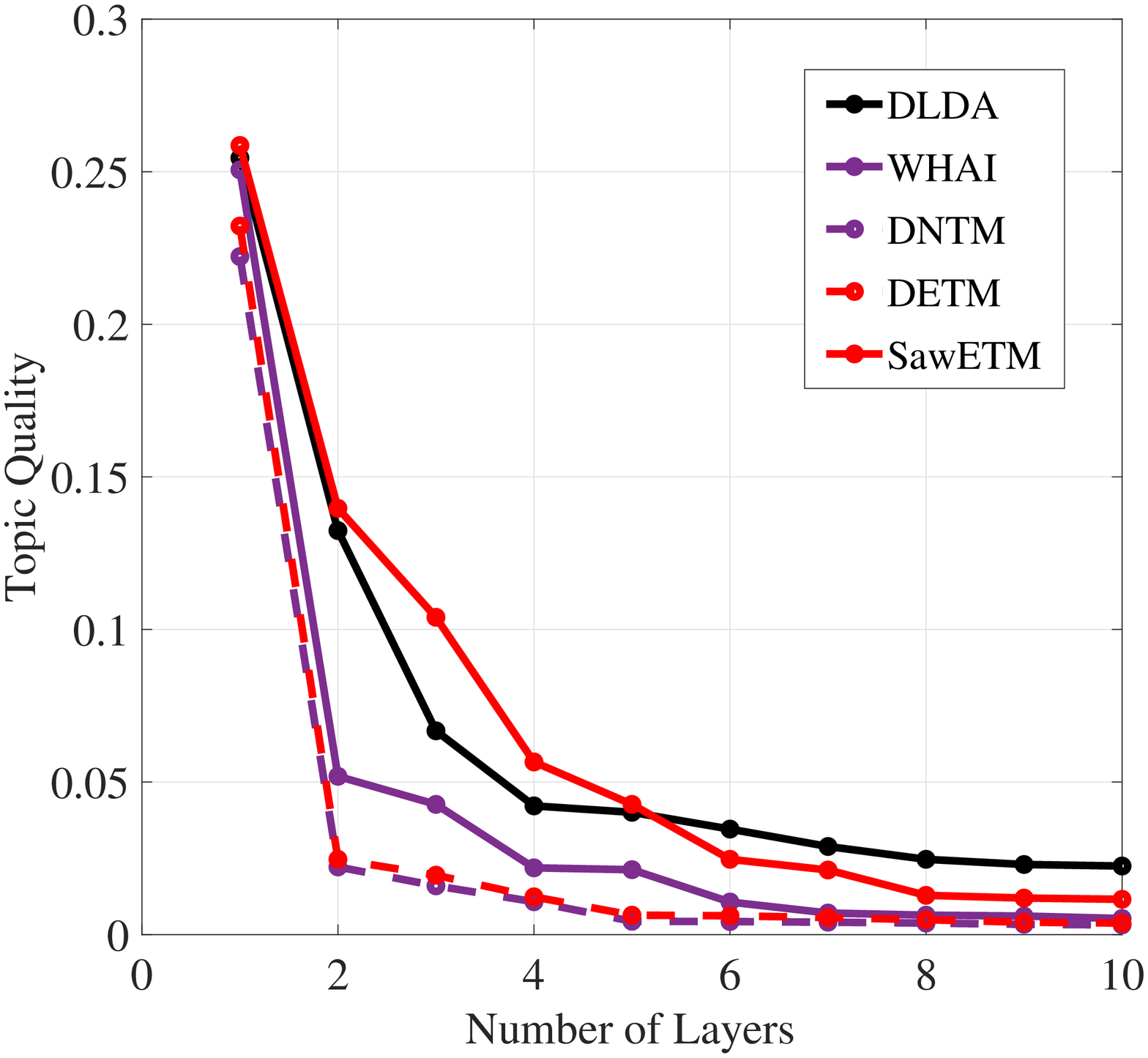}
} \hspace{-5mm}
\quad 
\subfigure[RCV1]{
\includegraphics[width=0.32\linewidth]{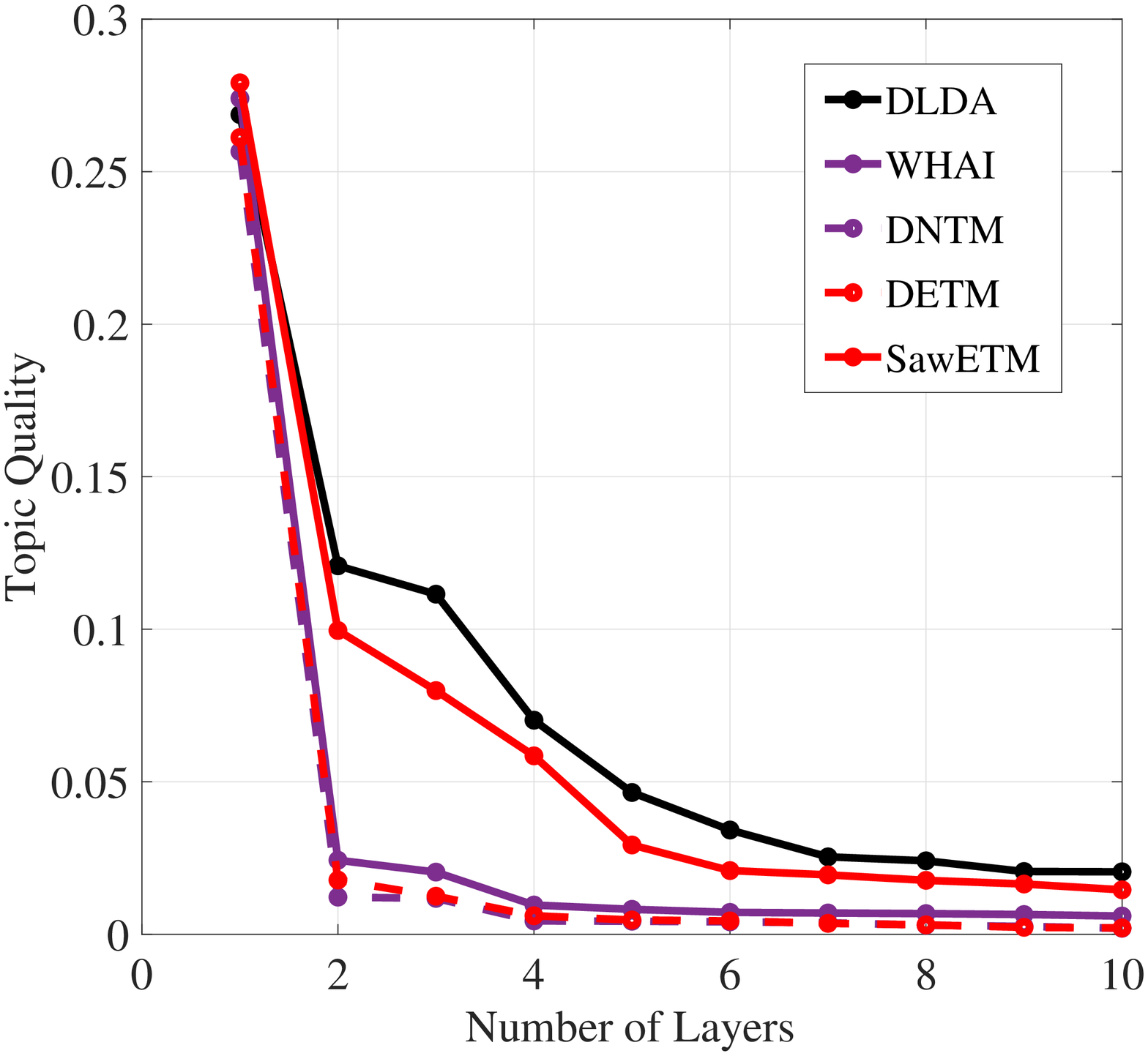}
} \hspace{-5mm}
\quad 
\subfigure[PG19]{
\includegraphics[width=0.32\linewidth]{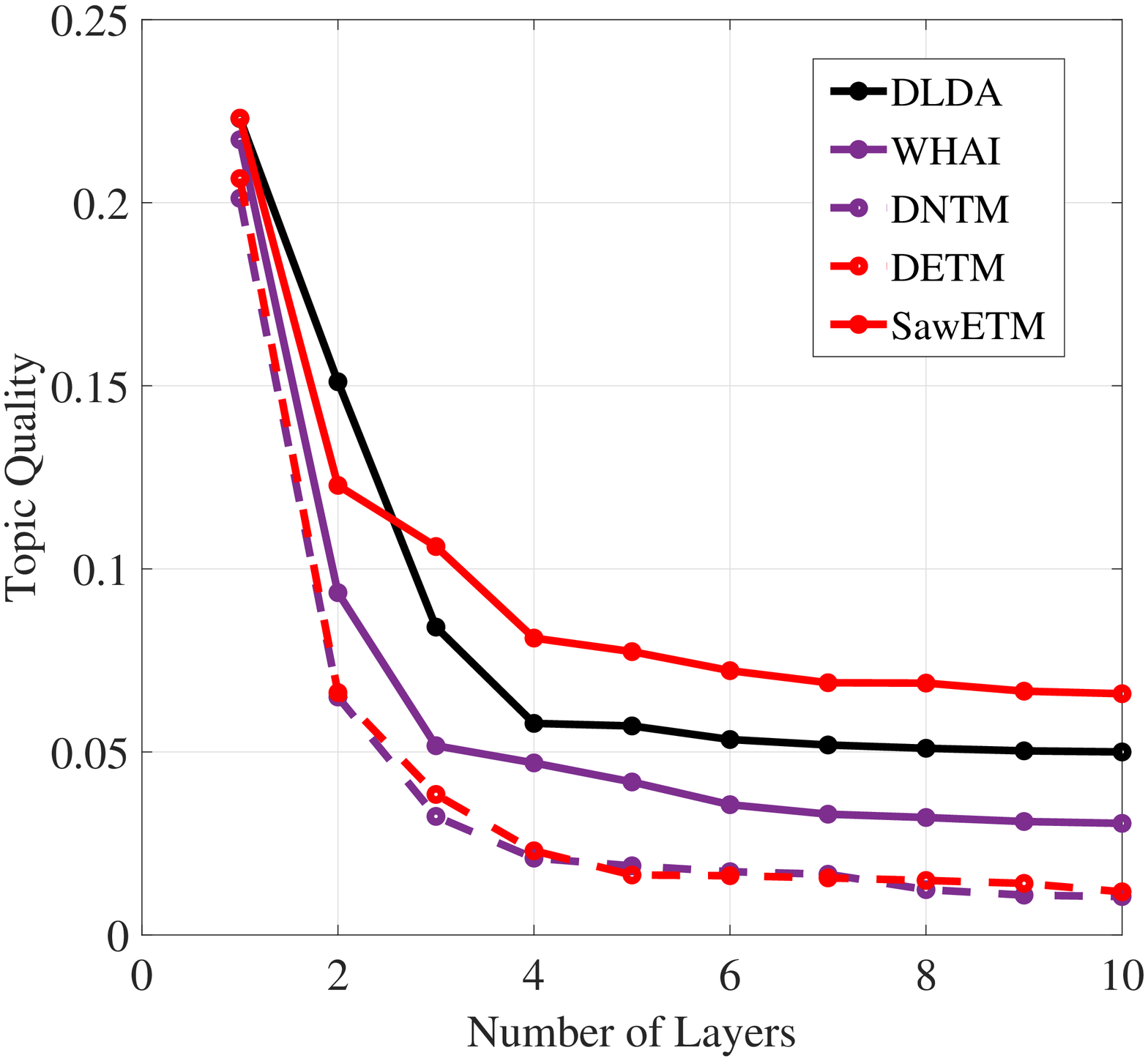}
}
\caption{(a)-(c): Comparison of per-heldout-word perplexity (the lower the better).
(d)-(f): Comparison of topic quality
(the higher the better).}
\label{figure:PPl}
\end{figure*} \vspace{-3mm}
\section{Experiments}
In this paper, SawETM is proposed for extracting deep latent features and analyzing documents unsupervised. So we evaluate the effectiveness of the proposed model by unsupervised learning tasks in this section. Specifically, four widely-used performance measure of topic models are used in the experiments, which include perplexity, topic quality, and document clustering accuracy/normalized mutual information metric. The experiments are conducted on four real-world datasets including both regular and big corpora. In order to further understand the proposed model and verify our motivation, we visually inspect the learned word and topic embeddings as well as topic hierarchies.  Our code is available at \url{https://github.com/BoChenGroup/SawETM}.


\subsection{Experimental settings}


\textbf{Datasets. } We run our experiments on four widely used benchmark corpora including \textbf{R8}, 20Newsgroups (\textbf{20NG}), Reuters Corpus
Volume I (\textbf{RCV1}), and \textbf{PG-19}. 
The R8 dataset is a subset of the Reuters 21578 dataset, which consists of $7,674$ documents from 8 different review groups. R8 is partitioned into a training set of $5,485$ ones and a testing set of $2,189$ ones.
The 20NG dataset, with a vocabulary size $36,534$, consists of $18,774$ documents from $20$ different news groups and its average document length is $221$. 20NG is split into a training set of $11,314$ ones and a testing set of $7,532$ one. 
The RCV1 dataset, with a vocabulary size $50,000$, consists of $804,414$ documents \cite{lewis2004rcv1} and its average dcoument length is $140$.
The PG-19 dataset is extracted from Project Gutenberg \cite{raecompressive2019} and contains $28,752$ book. We first build a vocabulary with $20,000$ words from this dataset, and then split each book with $1024$ tokens, which result in $613,386$ documents. 
For data processing, we first preprocess all the datasets by cleaning and tokenizing text, then removed stop words and low frequency words appearing less than 5 times, and finally select the N most frequent words to build a vocabulary. Note that, the R8 and 20NG datasets are used for document clustering experiments as there are ground-truth labels. 



\textbf{Baselines.} We compare SawETM with basic Bayesian topic models and neural topic models: 
1. LDA Group, single layer topic models, including Latent Dirichlet Allocation (\textbf{LDA}) \cite{blei2003latent}, which is a basic probability topic model; 
LDA with Products of Experts (\textbf{AVITM}) \cite{srivastava2017autoencoding}, which replaces the mixture model in LDA with a product of experts and uses the variational inference update parameters; 
LDA with word embeddings (\textbf{ETM}) \cite{dieng2020topic}, a generative model of documents that marries traditional topic models with word embeddings. 
2. DLDA Group, hierarchical topic models, including Deep latent allocation inferred by Gibbs sampling (\textbf{DLDA-Gibbs}) \cite{zhou2015poisson} and by TLASGR-MCMC (\textbf{DLDA-TLASGR}) \cite{cong2017deep};
and Weibull hybrid autoencoding inference model for Deep latent allocation (\textbf{WHAI}) \citep{zhang2018whai},
which employs a deep variational encoder to infer hierarchical document representations and updates the parameters by a hybrid of stochastic-gradient MCMC and variational inference. 
3. Variants Group, variants of SawETM, including deep neural topic model (\textbf{DNTM}) as introduced in Eq.~\ref{eq_w}, which directly models the dependence via the learnable parameters $W$; and deep embedding topic models (\textbf{DETM}) as introduced in Eq.~\ref{eq_unshare}, which employs a similar decomposition form as SC but does not share topic embeddings between two adjacent layers. These two Variant models use the same encoder with SawETM. 
For all the baseline models, we use their official default parameters with best-reported settings.

\textbf{Setting.} The hyperparameter settings used for the GBN group are similar to the ones used in \citet{zhang2018whai}. For the hierarchical topic models, the network structures of 15-layer models are set as [256, 224, 192, 160, 128, 112, 96, 80, 64, 56, 48, 40, 32, 16, 8]. For the embedding-based topic models such as ETM, DETM, and SawETM, we set the embedding size as 100. For the NTMs, we set the hidden size as 256. For optimization, the Adam optimizer \citep{kingma2014adam} is utilized with a learning rate of $1\text{e} ^{-2}$. The mini-batch size is set as 200 in all experiments. All experiments are performed on Nvidia GTX 8000 GPU and coded with PyTorch. 


\subsection{Per-heldout-word perplexity} \label{sec_ppl}
Per-heldout-word perplexity is a widely-used performance measure of topic models. Similar to \citet{zhou2015poisson}, for each corpus, we randomly
select 80\% of the word token from each document to form a training matrix ${\rm{T}}$, holding out the remaining 20\% to form a testing matrix ${\rm{Y}}$. We use ${\rm{T}}$ to train the model and calculate the per-held-word perplexity as
 $$\exp \left\{ { - {1 \over {{y_{..}}}}\sum\limits_{v = 1}^V {\sum\limits_{n = 1}^N {{y_{vn}}\ln {{\sum\nolimits_{s = 1}^S {\sum\nolimits_{k = 1}^{{K^1}} {\phi _{vk}^{\left( 1 \right)s}\theta _{kn}^{\left( 1 \right)s}} } } \over {\sum\nolimits_{s = 1}^S {\sum\nolimits_{v = 1}^V {\sum\nolimits_{k = 1}^{{K^1}} {\phi _{vk}^{\left( 1 \right)s}\theta _{kn}^{\left( 1 \right)s}} } } }}} } } \right\},$$
 
where $S$ is the total number of collected samples and ${y_{ \cdot  \cdot }} = \sum\nolimits_{v = 1}^V {\sum\nolimits_{n = 1}^N {{y_{vn}}} }$. 

Fig.~\ref{figure:PPl} (a)-(c) show how the perplexity changes as a function of the number of layers for various models over three different datasets. For both RCV1 and PG19, which are too large to run Gibbs sampling, we omit DLDA-Gibbs and only include DLDA-TLASGR for comparison. 
In the LDA group, ETM gets the best performance compared with LDA and AVITM, which can be attributed to the powerful word embeddings decoder
\cite{dieng2020topic}. LDA has the better performance compared with AVITM, which is not surprising as this batch algorithm can sample from the true posteriors given enough Gibbs sampling iterations. But these models are limited to single-layer shallow models, which can't benefit from the deep structure and results in the gap with the second group of models. 

Among these DLDA based models in group two, the DLDA-Gibbs outperforms other models, attributed to the more accurate posterior estimations, while DLDA-TLASGR is a mini-batch algorithm and slightly degraded performance in out-of-sample prediction. As a hierarchical neural topic model, WHAI gets the worse performance compared with DLDA-Gibbs/DLDA-TLASGR. Meanwhile, we can see that WHAI with a single hidden layer clearly outperforms AVITM, indicating that using the Weibull distribution is more appropriate than using the logistic normal distribution to model the document latent representation. Besides, the performance of DLDA-Gibbs/DLDA- TLASGR can be effectively improved by increasing the number of layers, while WHAI fails to improve its performance when the layer size becomes greater than three. This maybe due to all layers of the natural topic models are trained by SGD together, which makes it difficult to learn meaningful prior \cite{wu2021greedy}. And similar to deep VAE, this phenomenon is called posterior collapse \cite{sonderby2016ladder, maaloe2019biva}. 

With the powerful word embedding decoder and effective Weibull
upward-downward variational encoder, DETM of the variants group gets significant performance improvement with single layer. 
However, it also experience the similar problem with WHAI that no clear performance improvement when the number of layers becomes greater than three. 
Benefiting from the SC module between different layers, the learned knowledge at lower layers can flow to the upper layer, which can help the higher layers learn meaningful topics, resulting in better prior learned by SawETM. We can see that SawETM further improve the performance with the layer size becomes bigger, and get comparable performance with DLDA-Gibbs/DLDA-TLASGR.
Note that, although the improvement of SawETM is not that significant compared with DLDA-Gibbs/DLDA-TLASGR, DLDA-Gibbs/DLDA-TLASGR require iterative sampling to infer latent document representations in the testing stage, while SawETM can infer latent representations via direct projection, which makes it both scalable to large corpora and fast in out-of-sample prediction. Besides, thanks to SC and the improved encoder, SawETM can significantly outperform other NTMs.





\subsection{Topic quality}
A good topic model  can provide interpretable topics. In this section, we measure the model's performance in terms of topic interpretable \cite{dieng2020topic}. Specifically, topic coherence and topic diversity are combined here to evaluate topic interpretable/quality. Topic coherence is obtained by taking the average Normalized Pointwise Mutual Information (NPMI) of the top 20 words of each topic \cite{aletras2013evaluating}. It provides a quantitative measure of the interpretability of a topic \cite{mimno2011optimizing}. The second metric is topic diversity \cite{dieng2020topic}, which denotes the percentage of unique words in the top 20 words of all topics. Diversity close to 1 means more 
diverse topics. Topic quality is defined as the product between topic coherence and diversity.

Fig.~\ref{figure:PPl} (d)-(f) show the topic quality results of different layer for various models over three different datasets. Clearly, DLDA performs the best in terms of topic quality especially the higher layer, which is not surprising as all its parameters are updated by Gibbs sampling/TLASGR-MCMC \cite{cong2017deep}. 
Thanks to the use TLASGR-MCMC rather than a simple SGD procedure, WHAI consistently outperforms DNTM and DETM, which update all the parameters by SGD. Although equipped with a powerful word embeddings decoder, the topic quality of DETM clearly decreases as the number of layers increases. Through the above experimental phenomenon, we can find that it is difficult for NTMs to learn meaningful hierarchical topics. This is probably because NTMs often suffer from the posterior collapse problem in VAEs, making it hard to learn deeper semantic structure. 
However, SawETM achieving comparable performance to DLDA, which clearly outperforms the other deep neural topic models. As discussed in the Sec.~\ref{sec_ppl}, this improvement come from the SC module. The results of topic quality also agree with the results of perplexity, which are shown in Fig.~\ref{figure:PPl} (a)-(c).
\begin{table}[h]
	\centering
	\caption{Results of AC and NMI for document clustering task.}
    \label{Table:cluster}
    \scalebox{0.9}{
	\begin{tabular}{c|c|c|c|c|c}
    \toprule
     \multirow{2}{*}{\bf{Model}} & \multirow{2}{*}{\bf{Layer}} & \multicolumn{2}{|c|}{\bf{20News}} & \multicolumn{2}{c}{\bf{R8}}\\
      \cmidrule{3-6}& & AC & NMI & AC & NMI\\
        \midrule
		LDA & 1 & 46.52 & 45.15 & 51.41 & 40.47\\
        AVITM & 1 & 48.31 &46.33 & 52.43 & 41.20\\
        ETM & 1 &49.79 & 48.40 & 55.34 & 41.28\\
        \midrule
        PGBN & 1 & 46.62&45.43& 51.67& 40.76\\
        PGBN & 5 & 48.33&46.51 &54.21 & 41.21\\
        WHAI & 1& 49.43 & 46.56 & 57.86 & 42.31\\
        WHAI & 5 & 49.51 & 46.98 & 60.45& 43.98\\
        \midrule
        DNTM &1 &49.17 & 46.32 & 57.58 & 42.12 \\
        DNTM &5 & 49.25 & 46.79 & 59.93 & 43.90\\
        DETM &1 &50.24 & 48.69 & 61.21 & 43.45\\
        DETM &5  & 50.33 & 48.87 & 61.86 & 44.12\\
        SawETM&5 & \textbf{51.25} & \textbf{50.77} & \textbf{63.82} & \textbf{45.90}\\
	\bottomrule
	\end{tabular}} 
\end{table} \vspace{-3mm}
\begin{figure*}[!ht]
\centering
\subfigure[Word Embeddings]{
\includegraphics[width=0.35\linewidth]{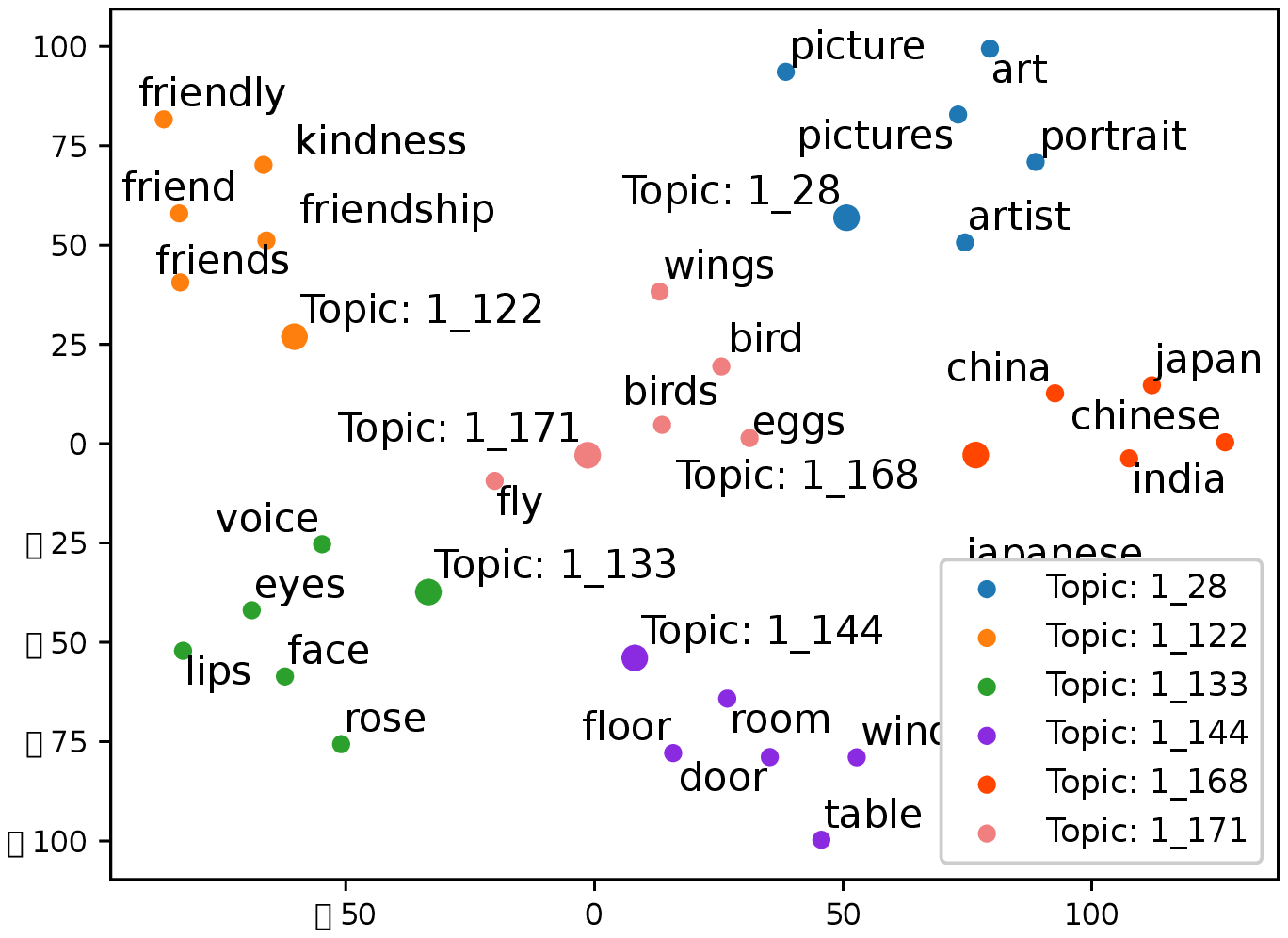}
\label{figure:word_embedding}
}
\quad
\subfigure[Topic Embedding]{
\includegraphics[width=0.60\linewidth]{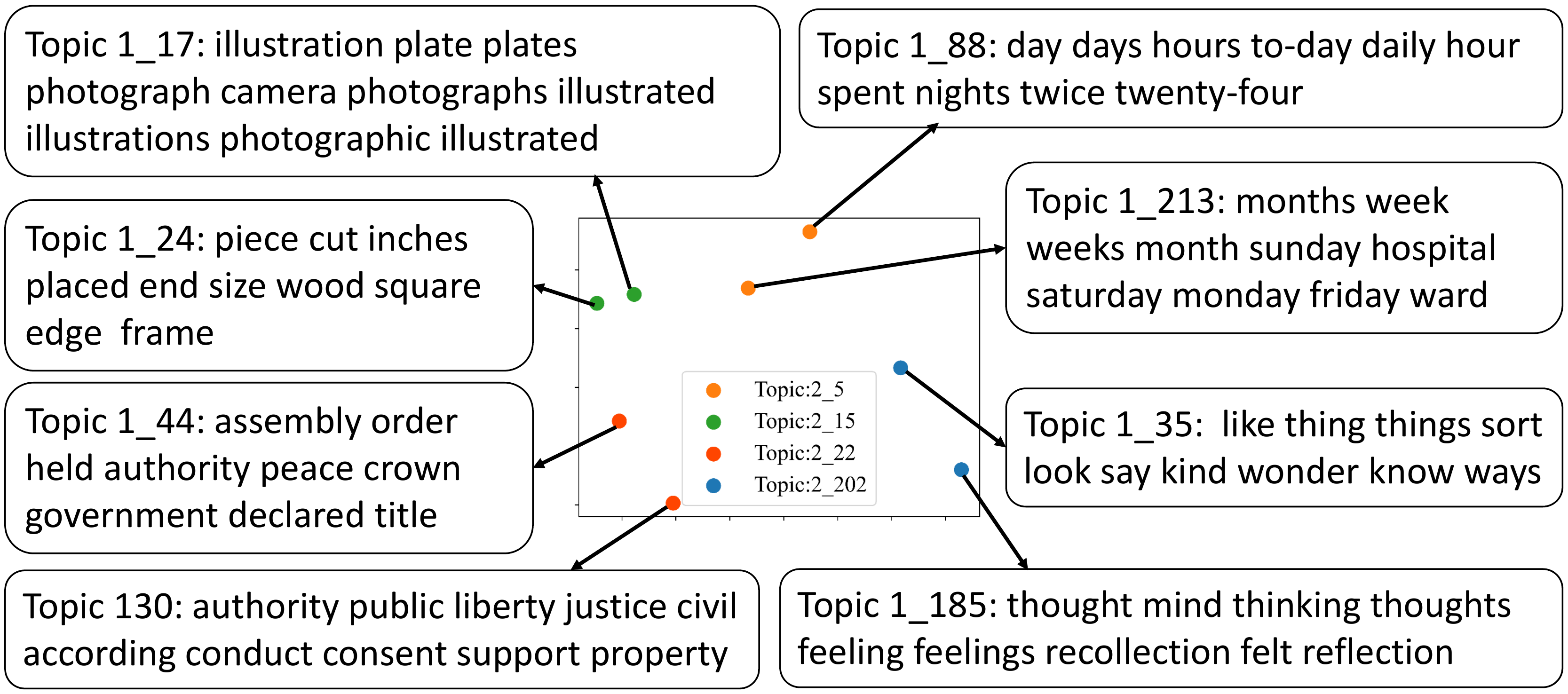}
\label{figure:topic_embedding}
} \vspace{-3mm}
\caption{t-SNE visualisation of (a) word embeddings, which we choose the top ten words for each topic at layer one and (b) topic embeddings, which we choose the top two sub topics for each topic at layer two. (Note that the Topic: $t\,\_\,j$ denotes the $j^{\text{th}}$ topic at $t^{\text{th}}$ layer.)} \label{figure:embedding}
\end{figure*}

\subsection{Document clustering}
We consider the multi-class classification task for predicting the clusters for test documents to evaluate the quality of latent document representations extracted by these models. In detail, we use the trained topic models to extract the latent representations of the testing documents and then use k-means to predict the clusters. As shown in Table \ref{Table:cluster}, the accuracy (AC) and  normalized mutual information metric (NMI) are used to measure the clustering performance, both of which are the higher the better. 
Tab.~\ref{Table:cluster} shows the clustering results of all the models on 20NG and R8 dataset.
It can be observed that with powerful word embeddings decoder and the Sawtooth Connection, SawETM can extract more expressive document latent representations and outperforms the other models included for comparison.
\begin{figure*}[!h]
\centering
\includegraphics[width=16cm]{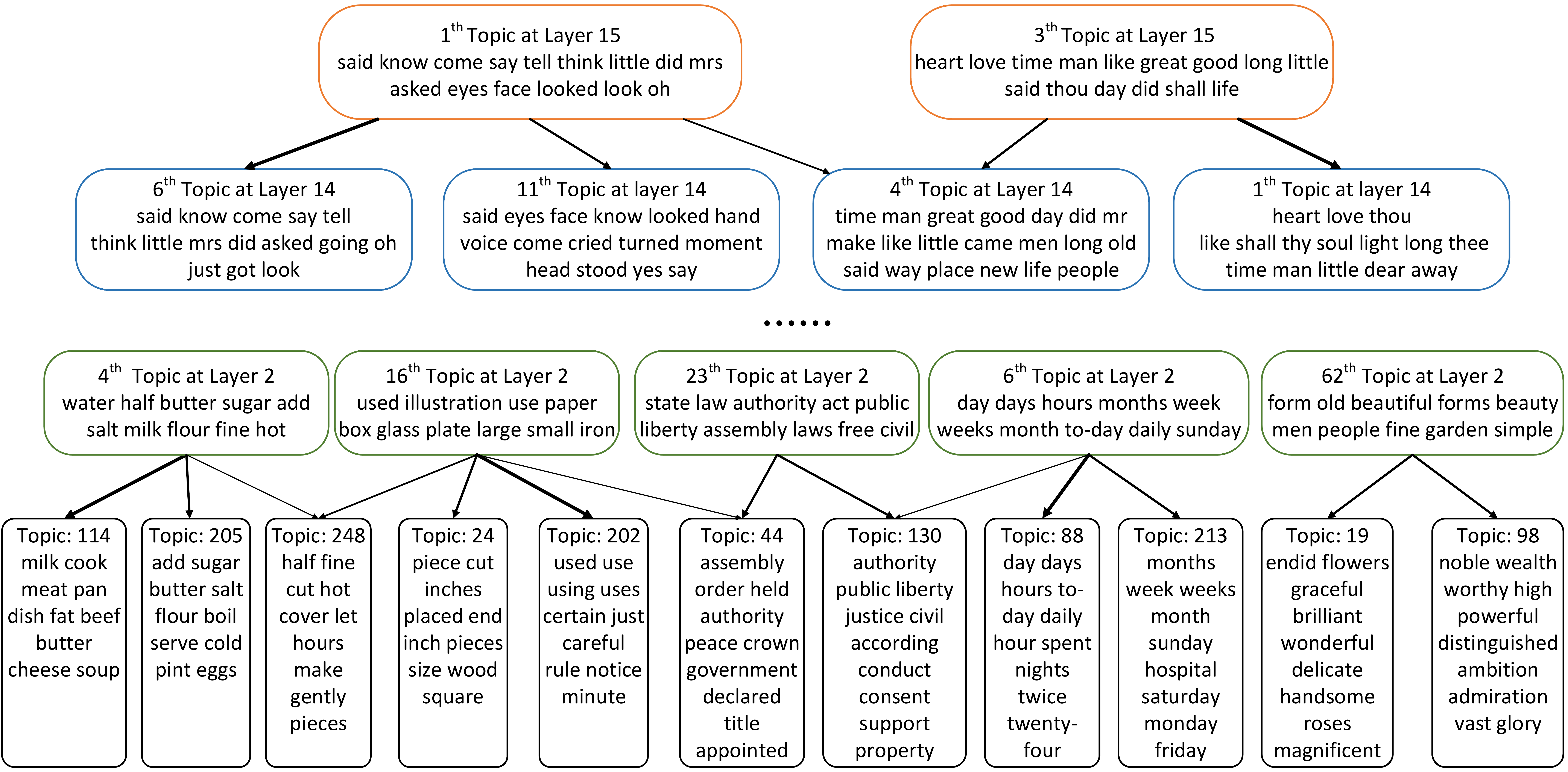}
\caption{An example of hierarchical topics learned from PG-19 by a 15-layer SawETM, We only show example topics at the top two layers and  bottom two layers.} 
\label{fig:hier_structure}\vspace{-2mm}
\end{figure*}\vspace{-2mm}
\subsection{Qualitative analysis}
One of the most appealing properties of SawETM is interpretability, we can visually inspect the inferred topics at different layers and the inferred connection weights between the topics of adjacent layers. Specifically, we conduct an extensive qualitative evaluation on the quality of the topics discovered by SawETM, including word embedding, topic embedding, and topic hierarchies. In this section, we use a 15-layer SawETM trained on  PG-19 for the visualization of embedding space and hierarchical structure experiments.\vspace{-3mm}
\paragraph{Visualisation of embedding space} The top 10 words from six topics are visualized in Fig.~\ref{figure:word_embedding} by t-SNE visualization \cite{van2008visualizing}. 
As we can see, the words under the same topic stay closer together, and the words under different topics are far apart. 
Besides, the words under the same topic are semantically more similar, which can demonstrate the meaning of the learned word embeddings.
Apart from the visualization of word embeddings, we visualize the topic embeddings. 
As shown in Fig.~\ref{figure:topic_embedding}, we select the top 2 sub-topics of the topic at the second layer. We can see that the sub-topic from the same topic have semantic similarity, and are closer in the embedding space. The above experiments show that the proposed SawETM learns not only meaningful word embeddings but also meaningful topic embeddings. More importantly, the learned words and hierarchical topics embedding can be projected into the same space, which can further support our motivations. \vspace{-3mm}
\paragraph{Hierarchical structure of topic model:}
Limited by the paper space, we only visualize topics at the top two layers and the bottom two layers. As shown in Fig.~\ref{fig:hier_structure}, the semantic meaning of each topic and the connections between the topics of adjacent layers are highly interpretable. In particular, SawETM can learn meaningful hierarchical topics at higher layers, indicating that it is able to support a deep structure.  \vspace{-2mm}
\section{Conclusion}
In this paper, we propose SawETM, a deep neural topic model that captures the dependencies and semantic similarities between the topics at different layers. We design a skip-connected upward-downward inference network to approximate the posterior distribution of a document.
Note that with the Sawtooth Connection technique, SawETM provides different views to a deep topic model, and further improves the performance of the neural deep topic model. 
As a fully variational deep topic model, SawETM can be optimized by SGD. Extensive experiments have shown that SawETM achieves comparable performance on perplexity, document clustering, and topic quality with the start-of-the-art model. In addition, with learned word and topic embeddings, and topic hierarchies, SawETM can discover interpretable structured topics, which helps to gain a better understanding of text data. \vspace{-2mm}
\section*{Acknowledgments}
Bo Chen acknowledges the support of NSFC (61771361), Shaanxi Youth Innovation Team Project, the 111 Project (No. B18039) and  the Program for Oversea Talent by Chinese Central Government. 

\bibliography{example_paper}
\bibliographystyle{icml2021}






\end{document}